\documentclass{elsart}
\usepackage{psfig,epsfig,subfigure}
\renewcommand{\thefootnote}{\fnsymbol{footnote}}

\newcommand{\gtsim}{\stackrel{>}{\sim}}

\begin{document}

\begin{frontmatter}
\title{Coherent $\pi^{\circ}$ photoproduction from
$^4$He\thanksref{dfg}}
\renewcommand{\thefootnote}{\arabic{footnote}}
\author[Goett]{F. Rambo\thanksref{rambdiss}}, 
\author[Mainz]{P. Achenbach},
\author[Mainz]{J. Ahrens}, 
\author[Mainz]{H.-J. Arends}, 
\author[Mainz]{R. Beck}, 
\author[Glasg]{S. J. Hall},
\author[Giess]{V. Hejny}, 
\author[Mainz]{P. Jennewein}, 
\author[MaiTa]{S.S. Kamalov\thanksref{kamaaddr}},
\author[Giess]{M. Kotulla}, 
\author[Giess]{B. Krusche}, 
\author[Goett]{V. Kuhr}, 
\author[Mainz]{R. Leukel}, 
\author[Giess]{V. Metag}, 
\author[Giess]{R. Novotny}, 
\author[Mainz]{V. Olmos de L\'eon}, 
\author[Mainz]{A. Schmidt},
\author[Goett]{M. Schumacher\thanksref{schuaddr}},
\author[Tuebi]{U. Siodlaczek}, 
\author[Goett]{F. Smend}, 
\author[Mainz]{H.~Str\"oher\thanksref{stroaddr}},
\author[Giess]{J. Weiss}, 
\author[Goett]{F. Wissmann}, 
\author[Giess]{M. Wolf}
\address[Goett]{II. Physikalisches Institut, Universit\"at 
G\"ottingen, Bunsenstrasse 7-9, D-37073 G\"ottingen, Germany}
\address[Mainz]{Institut f\"ur Kernphysik, Universit\"at Mainz, 
D-55099 Mainz, Germany} 
\address[Glasg]{Department of Physics and Astronomy, University of 
Glasgow, Glasgow G128QQ,  UK}
\address[Giess]{II. Physikalisches Institut, Universit\"at Giessen, 
D-35392 Giessen, Germany}
\address[MaiTa]{Institut f\"ur Kernphysik, Universit\"at Mainz, 
D-55099 Mainz, Germany and Department of Physics, National Taiwan 
University, Taipei 10617, Taiwan}
\address[Tuebi]{Physikalisches Institut, Universit\"at T\"ubingen, 
D-72076 T\"ubingen, Germany}

\thanks[dfg]{supported by Deutsche Forschungsgemeinschaft (SFB 201)}
\thanks[rambdiss]{Part of a doctoral thesis}
\thanks[kamaaddr]{Permanent address: Laboratory of Theoretical
Physics, JINR Dubna, Head Post Office Box 79,
SU-101000 Moscow, Russia} 
\thanks[schuaddr]{e-mail: 
Martin.Schumacher@phys.uni-goettingen.de} 
\thanks[stroaddr]{Present address: Institut f\"ur Kernphysik,
  Forschungszentrum J\"ulich, D-52425 J\"ulich, Germany}


\begin{abstract}
Differential cross sections and beam asymmetries for coherent 
$\pi^{\circ}$ photoproduction from $^4$He in the $\Delta$ energy-range 
have been measured with high statistical and systematic precisions 
using both decay photons for identifying the process. The experiment 
was performed at the MAinz MIcrotron using the TAPS photon 
spectrometer and the Glasgow/Mainz tagged photon facility. 
The differential cross 
sections are in excellent agreement with predictions based on the DWIA 
if an appropriate parametrization of the $\Delta$-nuclear interaction 
is applied. The beam asymmetries are interpreted in
terms of degrees of linear polarization of collimated coherent 
bremsstrahlung. The expected increase of the degree of linear 
polarization with decreasing collimation angle is  confirmed. 
Agreement with calculations is obtained on a few-percent level of 
precision in the  maxima of the coherent peaks.

 
\end{abstract}

\end{frontmatter}

\section{Introduction}

Coherent photoproduction of $\pi^{\circ}$ mesons from nuclei is an 
ideal tool to study fundamental properties of pion-nucleus 
interactions. Another important aspect in the study of this reaction 
is the unique opportunity 
to get information about modifications of the 
$\Delta(1232)$-resonance characteristics in the nuclear medium.
The nucleus $^4$He is especially suited for studies of this type 
because it combines the  well known structure of a few-body system 
with the large binding energy of a complex nucleus. 
Furthermore, because of the high excitation energy of 20 MeV of the 
first excited level it is comparatively easy to separate coherent 
from incoherent $\pi^{\circ}$ photoproduction processes.
The advantage of coherent  $\pi^{\circ}$ photoproduction as compared 
to, e.g., the production of charged mesons
is that the nucleus is identical in the initial and final states. 
This eliminates complicated nuclear rearrangement processes and 
makes a clear-cut theoretical interpretation possible. 
Furthermore,
the coherent $\pi^{\circ}$ photoproduction process is uniformly 
distributed over the whole nuclear 
volume. This leads to a substantial enhancement of the cross section 
in contrast to, e.g., $\pi^+$ photoproduction.

In spite of the fundamental
importance of this process the available experimental data were very 
scarce. There were only four papers available 
\cite{Sta69,Lef70,Tieger84,Ananin85} providing few and partly 
controversial
differential cross sections. The reason for this rather unsatisfactory
experimental situation was that the experiments were extremely 
difficult with the previously available photon beams and photon 
detectors. A major
step forward came with the advent of $cw$ electron accelerators 
\cite{Her76,Her81,Her90}
and of broad-band tagging facilities \cite{Ant91,hall96}
as they are now available, e.g., at MAMI in Mainz. A further
large  progress came with the availability of high energy-resolution
and large angular-acceptance  photon detectors.

The present experiment is a continuation of our previous research
on coherent
$\pi^\circ$ photoproduction on $^4$He \cite{kraus97} performed at 
MAMI (Mainz) \cite{Her76,Her81,Her90,Ant91,hall96}
which were mainly 
carried out to measure the degree of linear polarization of coherent 
bremsstrahlung  produced via 
a diamond radiator \cite{lohmann94}. In these previous investigations 
\cite{kraus97}
the 48 cm \O $\times$ 64 cm Mainz NaI(Tl) detector \cite{wiss94}
has been used as
a high detection-efficiency and high energy-resolution (1.5\%) 
spectrometer
for the high-energy photon emitted in an asymmetric $\pi^\circ$ 
decay. The experiment \cite{kraus97}
had been carried out for two different
collimation angles   (half of the angular opening)
of $\theta_{coll}$=0.7 mrad and $\theta_{coll}$=0.5 
mrad of the photon beam, corresponding to 
$\theta_{coll}=1.17 \times \theta_{BH}$ and 
$\theta_{coll}=0.83 \times \theta_{BH}$ where  $\theta_{BH}$=mc$^2$/E 
(m, E = electron mass and energy, respectively) is 
the characteristic   (Bethe-Heitler) angle. In this range
of collimation angles a pronounced dependence of the degree of linear
polarization on the collimation angle $\theta_{coll}$ is expected and 
this effect was clearly observed for the first time in that experiment 
\cite{kraus97}. 
In parallel to the experiment  a program was started 
to  precisely  calculate the degree of linear polarization including 
all geometrical effects like the parameters of the electron beam and 
the collimation of the photon beam \cite{rambo98}.  As a result of 
these investigations \cite{kraus97,rambo98} an agreement between  
measured and calculated degrees
of linear polarization on a $\gtsim$ 3\% level of precision 
was obtained.

Differing from our previous experiment \cite{kraus97} where one
decay photon has been used to identify a $\pi^{\circ}$ event, our 
present experiment carried out with the TAPS setup 
\cite{nov87,nov91,nov96} at MAMI
makes use of both decay photons and a much larger angular acceptance. 
Therefore, this latter method permits
to measure the degree of linear polarization of collimated photons
with higher systematic and statistical precisions than it was possible 
before \cite{kraus97,rambo98}. Furthermore, it was possible to evaluate
differential cross sections for coherent $\pi^{\circ}$ photoproduction 
in wide energy and angular ranges.

After completion of the present paper a similar experiment performed
by Bellini et al. \cite{belli99} came to our attention. It used the 
LEGS tagged polarized photon beam and covers a similar, but 
smaller range of photon energies with a smaller number of pion 
production angles. Using one large $48 cm \oslash \times 48 cm $ 
NaI(Tl) detector for the
high-energy photon and an array of smaller NaI(Tl) detectors for 
the low-energy photon of the asymmetric $\pi^\circ$ decay it was 
possible to single out the coherent 
${\vec \gamma} + ^4{}$He $\to \pi^\circ + ^4{}$He production channel.
The present and the previous data \cite{belli99} are in a general
reasonable agreement with each other except for some deviations especially
at  the highest energies
accessible in \cite{belli99}, as will be shown later.

\section{Experiment}

The experiments have been carried out using the BaF$_2$ multi-detector
array TAPS \cite{nov87,nov91,nov96} at the tagged-photon facility 
of the A2-collaboration \cite{Ant91,hall96} installed at the microtron 
MAMI in Mainz \cite{Her76,Her81,Her90}.  Using a diamond crystal as 
radiator and positioning a collimator at a distance of 2500 mm 
downstream
from the radiator the tagged-photon facility provides linearly 
polarized photons with degrees of linear polarization 
up to 50 $-$ 60\% \cite{kraus97}.
The accelerator was operated at its nominal  
maximum  energy of 855 MeV.
The intensity of the photon flux is limited by the tagging technique
which does not allow electron rates in a tagger channel ($\Delta$ E
$\approx$ 2 MeV) larger than
10$^6$ s$^{-1}$. Since the  
electron rates are strongly 
increasing with decreasing energies of the bremsstrahlung photons
the useful photon flux can be increased by switching off 
unused tagger channels at low photon energies. 
In the present experiment
the orientation   of the diamond crystal was  chosen such that
the position of the discontinuity at the high-energy side of the
coherent-bremsstrahlung  spectrum produced 
by the most prominent reciprocal lattice vector $[02\overline{2}]$ 
was located close to  360 MeV.  
This means that significant linear polarization appeared within the 
energy range between 200 MeV and 360 MeV. Therefore, the tagging
channels corresponding to photon energies below 200 MeV were 
switched off for the purpose mentioned above.

\begin{figure}
\centering\epsfig{figure=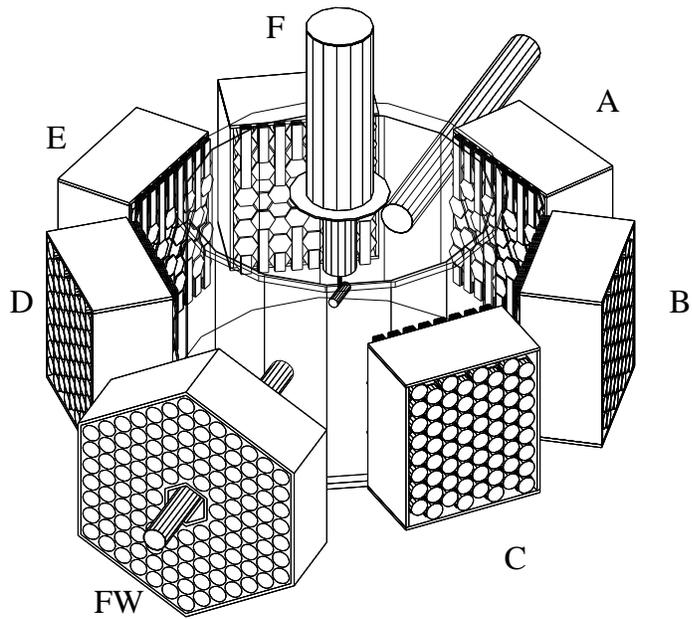,width=.7\linewidth}
\caption{The setup of the spectrometer TAPS as used for the present 
experiment. Six BaF$_2$ blocks A - F and the forward wall FW are 
arranged around the scattering chamber which is indicated by thin 
lines. In the center of the scattering chamber the 
$^4$He target is shown as a horizontal cylinder. The photon beam 
enters the scattering chamber through the beam pipe between blocks 
A and F and leaves it through a hole  in the forward wall FW.}
\label{fig1}
\end{figure}

The setup of the experiment is shown in Fig. \ref{fig1}.
TAPS consists of 384 BaF$_2$ modules having 25 cm length and hexagonal
cross section (inner diameter 5.9 cm), mounted in six blocks
of 64 \cite{Krusche95,hejny98}. 
Each module  is equipped with 
its individual veto detector in front to identify incoming charged 
particles. The blocks were arranged in a horizontal
plane in a distance of 57 cm from the target center. 
Their central axes were located  in the scattering plane at polar 
angles of
-149$^\circ$, -99$^\circ$, -49$^\circ$, 50$^\circ$, 100$^\circ$
and 151$^\circ$, respectively.  
Additional 120 BaF$_2$ modules were mounted in the forward direction
symmetrically  around the photon beam to build a  hexagonal forward 
wall (FW)
which covered a range of polar angles from 2$^\circ$ to 19$^\circ$. 
The FW modules are of the phoswich type with a  plastic detector 
mounted directly  onto the crystal. The light output from both parts 
is viewed by the same photomultiplier tube. This gives an excellent
identification of charged particles in addition to that via the 
pulse-shape discrimination provided by BaF$_2$.

The liquid helium target had a diameter of 3 cm and a length of 10 cm,
corresponding to about 1.86$\times$10$^{23}$\,nuclei/cm$^2$. The 
target cell was made of Kapton and
was mounted in a large (\O\,90\,cm), evacuated scattering chamber
of carbon fiber 
to minimize scattering, conversion and energy loss between target and 
detector. For automatic refilling, the Kapton cell was connected to a
5 liter $^4$He Dewar vessel contained in a lq. N$_2$ shield above 
the scattering  chamber.  Several layers of 
superisolation foil were used to protect the target from thermal 
radiation. With these precautions an uninterrupted measurement 
was possible for 9 h.

Three different collimators  having diameters of 3 mm, 4 mm and 6 mm 
were used to study the effect of collimation on the degree of linear 
polarization. In order to obtain approximately the same statistical 
precision in the three measurements, the total beam time of 125 h 
on the target was subdivided  into three periods of 57, 43 and 25 h, 
respectively.

\section{Data analysis and results}

The two-photon decay was used to identify neutral pions. In the 
evaluation of data,
events were selected which contained at least two coincident photons, 
being unambiguously
identified using pulse-shape analysis, information from the 
charged-particle veto detectors, and time-of-flight analysis.

As a first step to identify $\pi^\circ$ mesons from coincident 
(within a time window of 0.7 ns) photon pairs in two different
TAPS blocks,
spectra of invariant  masses $m_{\gamma\gamma}$ were constructed 
where the invariant mass is given by
\begin{equation}
m_{\gamma\gamma}=\sqrt{2E_1E_2(1-\cos\Phi)},
\label{meff}
\end{equation}
with $E_1$ and $E_2$ being the photon energies and $\Phi$ the angle 
between the two photons. Since most of the two-photon events are due 
to $\pi^\circ$ or $\eta$ decays, the spectrum of invariant masses 
contains two pronounced peaks in addition to a small combinatoric 
background from events with more than two photons in the time window.
These peaks are  slightly asymmetric because of the asymmetric
response function of the BaF$_2$ detectors \cite{Gabler93}. To take 
care of this,
also the window of accepted events around the invariant mass of 
135 MeV had to be chosen asymmetric, i. e.  from 90 to 160 MeV.

Because of  the high angular resolution of TAPS  the opening angle
$\Phi$ between two 
decay photons has been used to separate  coherent from incoherent 
$\pi^\circ$ photoproduction processes. For this purpose use was made
of the relation
\begin{equation}
\Phi_{min} = 2\arccos \left( \frac{\sqrt{(E_{\pi^{\circ}}^{lab})^2 
- m_{\pi^{\circ}}^2}}{E_{\pi^{\circ}}^{lab}}  \right)
\label{Phimin}
\end{equation}
between the minimum opening angle
$\Phi_{min}$ which corresponds to the symmetric two-photon decay
and the $\pi^{\circ}$ energy $E_{\pi^{\circ}}^{lab}$
in the laboratory. Since all incoherent processes
lead to a breakup of the helium nucleus, the  respective
$\pi^\circ$ mesons have at least 20 MeV
less energy than the coherently produced ones
for a given and, in the following discussion, fixed primary
photon energy. Correspondingly, the  opening angles $\Phi$
from incoherent processes are shifted to distinctly higher
values (typically by 3 times the angular resolution).
This shift defines, via Eq. (\ref{Phimin}), 
a range of opening angles $\Phi$ which only contains coherently 
produced $\pi^\circ$ mesons. 

\begin{figure}
\centering\epsfig{figure=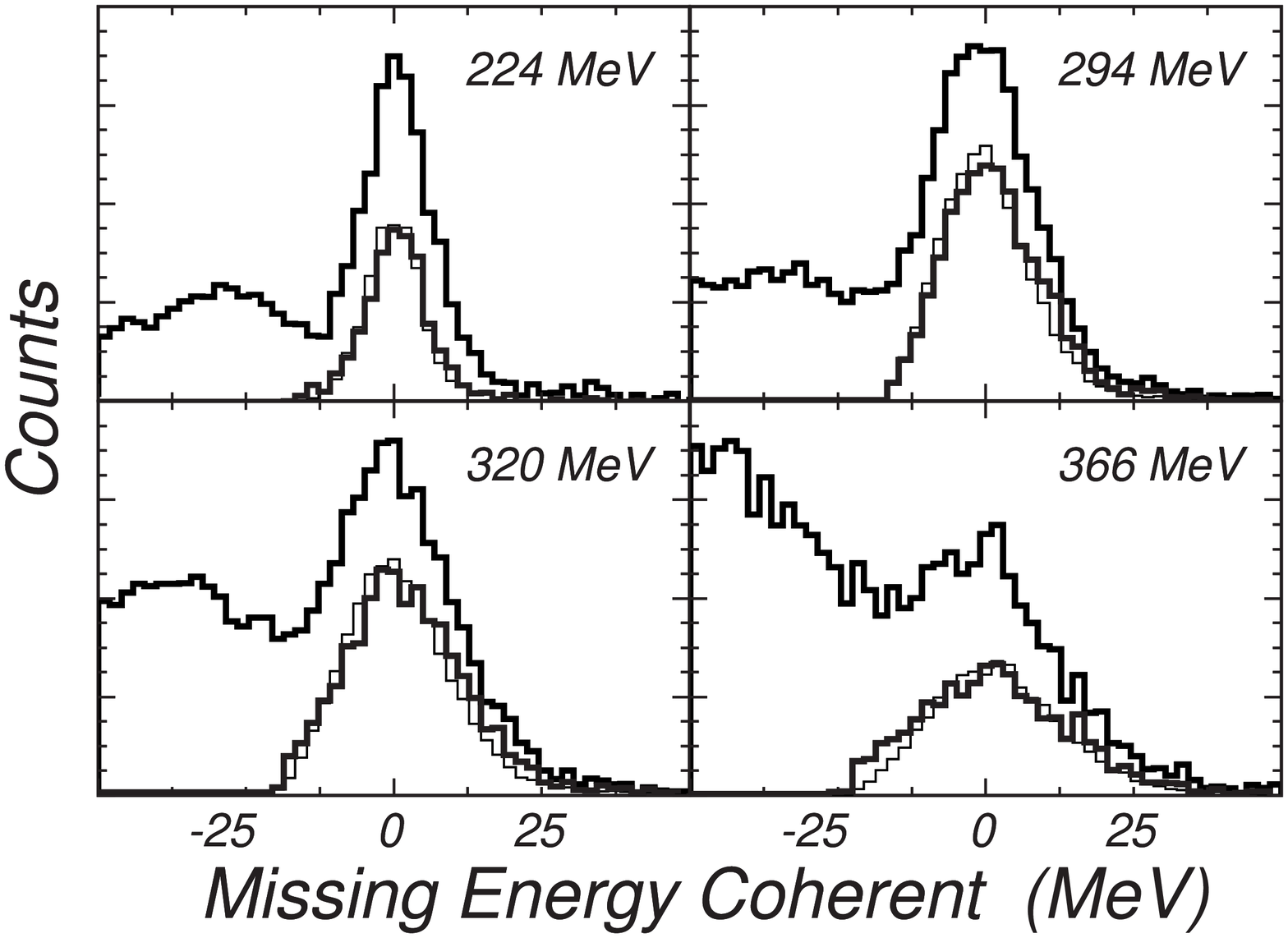,width=1.0\linewidth}
\caption{Numbers of events of $\pi^\circ$ photoproduction 
at primary photon energies of 224 MeV, 294 MeV,
320 MeV and 366 MeV versus 
the missing energy $\Delta E_{\pi^{\circ}}$ calculated via 
Eq. (\ref{missing}) for the coherent kinematics.
The upper distributions represented by thick solid lines
contain all $\pi^\circ$ events identified 
by the invariant mass defined in Eq. (\ref{meff})
of their decay photons. The lower distributions indicated 
by thick solid  lines represent the remaining
experimental events after the cut on the opening angle $\Phi$ of the 
decay photons is carried out. The thin solid lower lines show
the results of  Monte Carlo simulations in which only
coherently produced $\pi^\circ$ mesons were generated. } 
\label{fig2}
\end{figure}

\begin{figure}
\centering\psfig{figure=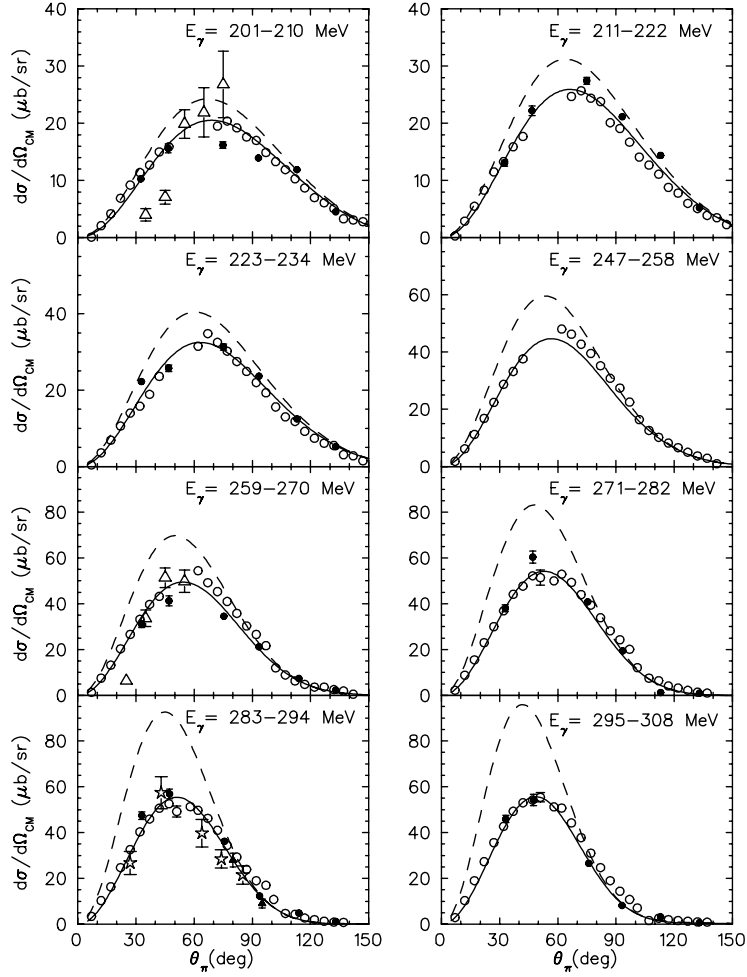,width=.7\linewidth}
\caption{Unpolarized differential cross sections 
for coherent $\pi^\circ$ photoproduction from $^4$He
for different photon energy intervals $E_\gamma$ as given in the figures. 
Open circles:  present experiment, 
full circles: BNL data ref. \cite{belli99},
open triangles: Tomsk data ref. \cite{Ananin85},
open stars: MIT data ref. \cite{Tieger84},
full triangles: ref. \cite{Lef70}.
Dashed curves are the DWIA results of ref. \cite{drechsel99}.
The solid curves are the theoretical results 
obtained with the formfactor-type parameterization for the $\Delta$ 
self-energy of ref. \cite{drechsel99}.}
\label{fig3}
\end{figure}

\begin{figure}
\centering\psfig{figure=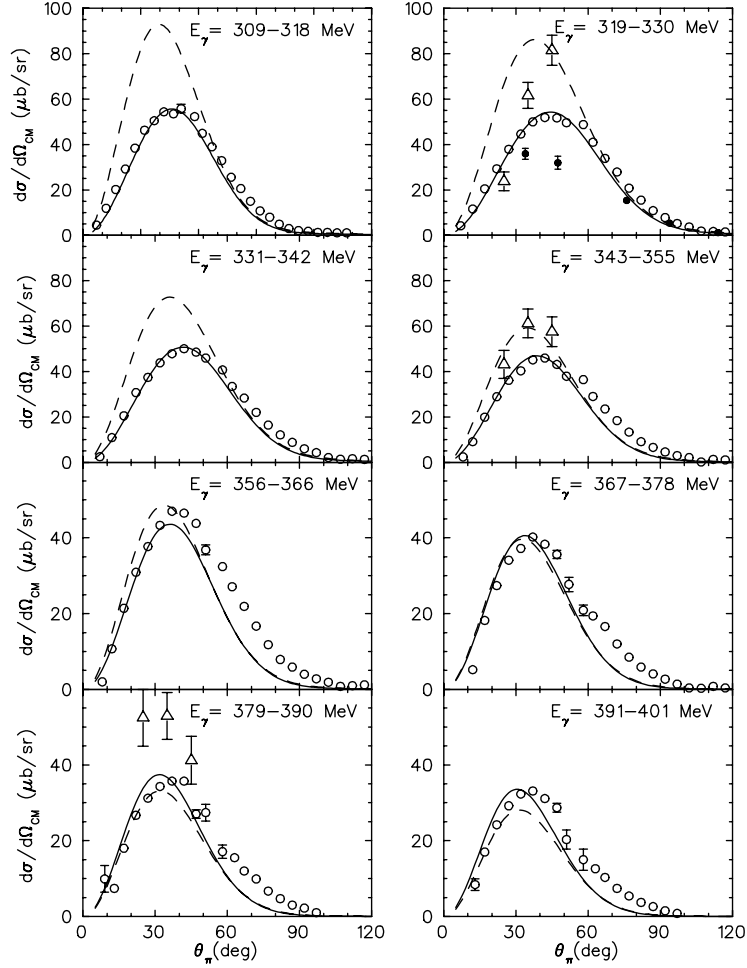,width=.7\linewidth}
\caption{Unpolarized differential cross sections 
for coherent $\pi^\circ$ photoproduction from $^4$He
for different photon energy-intervals $E_\gamma$ as given in the figures.
The notations are the same as in Figure \ref{fig3}.}
\label{fig4}
\end{figure}

For the verification of the effectiveness of the procedure of 
separating coherent from incoherent events described above, 
use has been made of an analysis of
the ``missing energy'' of the event which makes an explicit use
of the energy information provided by the tagger. Assuming coherent
$\pi^\circ$ photoproduction the $cm$ energy of the $\pi^\circ$ 
meson can be calculated via 
\begin{equation}
E^*_{\pi^{\circ}}(E_\gamma)=
\frac{s+m_{\pi^{\circ}}^2-m^2_{He}}{2\sqrt{s}}
\label{miss1}
\end{equation}
from the energy $E_{\gamma}$ of the primary photon, where 
${\sqrt s}=\sqrt{2E_\gamma m_{He}+m^2_{He}}$ is the total energy in the 
$cm$ system and $m_{He}$  the mass of the He nucleus. On the other 
hand the same energy 
can be obtained via Lorentz transformation from the measured
energies $E_1$ and $E_2$ and emission angles $\Theta_1$ and $\Theta_2$
of the decay photons in the laboratory system via
\begin{equation}
E^*_{\pi^{\circ}}(\gamma_1,\gamma_2)=
\gamma(E_1+E_2-\beta(E_1\cos\Theta_1+E_2\cos\Theta_2)),
\label{miss2}
\end{equation}
with $\beta=E_\gamma/(E_\gamma+m_{He})$ and $\gamma=\sqrt{1-\beta^2}$.  
The difference of the two energies
\begin{equation}
 \Delta E_{\pi^{\circ}} =
E^*_{\pi^{\circ}}(\gamma_1,\gamma_2)-E^*_{\pi^{\circ}}(E_\gamma)
\label{missing}
\end{equation}
is the ``missing energy'' which should be zero in case of coherent
$\pi^\circ$ photoproduction for an infinitely high precision of the
experimental quantities.
The verification of the effectiveness of the procedure 
of separating coherent from incoherent processes is illustrated in
Fig. \ref{fig2}, 
where numbers of events are shown as a function of the missing energy
$\Delta E_{\pi^{\circ}}$
for the four primary photon energies of $E_\gamma$= 224, 294, 320 and 
366 MeV. The upper curves contain the numbers of events without cuts 
in the spectra of opening angles $\Phi$ as described in the 
preceding paragraph. 
These curves contain a peak structure around zero missing energy 
which is mainly due to coherent $\pi^\circ$ photoproduction and a 
tail at the negative-energy side which is due to incoherent 
$\pi^{\circ}$ photoproduction.
The two lower curves are (i) experimental  numbers of events
after the cuts in the spectra of opening angles $\Phi$ are made,  
and  (ii) 
completely analogously simulated \cite{Geant94} numbers of events 
where the
simulation is carried out under the premise of coherent $\pi^\circ$
photoproduction.  These two lower curves are adjusted to each other 
to give the same area.
This, apparently, leads to a very good overall agreement of the
two distributions, thus  
proving that indeed the respective experimental data are due to
coherent $\pi^\circ$ photoproduction only. A  further quantity which 
is obtained from the  simulation  \cite{Geant94} 
is the detection efficiency  of the detector which
relates numbers of events to differential cross sections. Using this
detection efficiency the experimental differential cross sections shown in Figs.
\ref{fig3} and \ref{fig4} and in Table \ref{sigdif} are obtained. 
The errors given are statistical errors
only. Table \ref{sigtot} shows the total cross sections. 
The total systematic error of the cross section is estimated to be
7\%.
It stems from the following contributions:\\
(i)  The detection efficiency of the TAPS apparatus: 3\%, \\
(ii) the analysis cuts in the pulse-shape and time-of-flight 
spectra: 5 \%, \\ 
(iii) the tagging efficiency for coherent bremsstrahlung which is 
more strongly 
influenced by beam instabilities than the one for incoherent 
bremsstrahlung: 3\%, \\ 
(iv) dead-time corrections of the photon flux: 1.5 \% , and \\
(v) target thickness: 2\%.

As is illustrated by the reduction of the numbers of experimental
events shown in 
the peaks around zero missing  energy of Fig. \ref{fig2},
the cuts on the
opening angle $\Phi$  have the unavoidable side effect of decreasing
the acceptance of the TAPS detector. In the angular ranges 
between the TAPS blocks the acceptance almost drops to zero. 
This leads to gaps in the experimental differential cross sections
close to the maxima of the angular distributions, 
as can be seen in part of the experimental data shown in 
Figs. \ref{fig3} and \ref{fig4}. Fortunately, 
the differential cross
sections are smooth functions of polar angle so that the gaps may 
be safely bridged by interpolation.

The data analysis described so far averages over the two directions 
(vertical and horizontal) of
linear polarization so that differential cross sections for 
unpolarized photons are obtained. In a further step of analysis use 
has been made
of the linear polarization provided by coherent bremsstrahlung and 
of the properties of coherent $\pi^\circ$ photoproduction on $^4$He 
as a polarimeter for measuring the degree of linear polarization on 
an absolute scale. These latter properties are provided by the fact 
that both the nucleus and the $\pi^\circ$ meson have no spins, 
so that $\pi^\circ$
mesons are emitted exclusively as $p$ waves through $M1$ 
excitation of the $\Delta$ resonance,  a channel which is already 
strongly favored for the free nucleon. 
As a consequence, the degree of linear polarization of the photon 
beam is completely transferred 
to the azimuthal asymmetry of the $\pi^\circ$ mesons. 
In this case the general expression for the differential cross 
section in the $cm$ frame can be given by
\begin{equation}
\frac{d\sigma}{d\Omega}=| (\vec{\epsilon} \times 
\hat{\vec{k}})\cdot \hat{\vec{q}}|^2 |F(\vec{q},\vec{k})|^2,
\label{lin}
\end{equation}
where $\vec{\epsilon}$ denotes the polarization vector,  
$\hat{\vec{k}}$ the direction of the photon and $\hat{\vec{q}}$ 
the direction of the pion.
The function $F(\vec{q},\vec{k})$ is a polarization-independent 
normalization factor.

The azimuthal acceptance of TAPS is limited to angular intervals
of $\Delta\varphi^{max}=\pm 25^\circ$  width  around 
$\varphi=0^\circ$ and $\varphi=180^\circ$, i.e., above and below 
the horizontal reaction plane 
on both sides of the photon beam. In order to make use of the full
sizes of these intervals without losing information, 
the azimuthal distributions $N_V(\varphi)$ and  
$N_H(\varphi)$ of coherent $\pi^\circ$ events were determined 
separately for each tagger channel. These two quantities
$N_V(\varphi)$ and  $N_H(\varphi)$ are the  numbers of events from
coherent $\pi^\circ$ photoproduction as  functions of the azimuthal
angle $\varphi$ of the $\pi^\circ$ meson with the electric vector
 being preferentially vertical 
(vertical polarization V)
and horizontal  (horizontal polarization H), respectively, 
normalized to the same numbers of primary photons. The angular 
characteristic given by
Eq. (\ref{lin}) predicts that the numbers of events $N_V(\varphi)$ 
are larger than the numbers of events $N_H(\varphi)$ with the 
intensity ratio being determined by the degree of linear 
polarization (in case of complete
linear polarization $N_H(\varphi)$ would be zero).
The degree of linear polarization  was evaluated via
\begin{equation}
P(E_\gamma) = \frac{1}{\cos 2\varphi}
              \frac{N_V(\varphi) - N_H(\varphi)}{N_V(\varphi) + 
              N_H(\varphi)}~~
\label{phi-dep}
\end{equation}
using angular bins  of $\Delta\varphi^{bin}= 3^\circ$ widths.
As expected,   Eq. (\ref{phi-dep}) neither contains the
efficiency of the detector nor the absolute value of the photon
flux.  This fact  to a large extent removes the systematic errors  
discussed above
in connection with the differential cross sections. The factor
$1 / \cos{2\varphi}$ corrects for the azimuthal distribution of the
$\pi^\circ$ mesons and is equal to 1 in case the $\pi^\circ$ meson is 
emitted exactly in the horizontal plane.
The final result for the degree of linear polarization in each tagger 
channel is obtained by averaging the results obtained from the
expression on the r.h.s. of 
Eq. (\ref{phi-dep}) over the angular bins $\Delta\varphi^{bin}$
between  $\Delta\varphi^{max}=- 25^\circ$  and 
$\Delta\varphi^{max}=+ 25^\circ$.

\section{Discussion}

\subsection{Cross sections for coherent $\pi^{\circ}$ photoproduction}

The differential cross sections obtained in the present experiment
for  photon energies between 200 MeV and 400 MeV
are  compared with previous experimental data 
\cite{Sta69,Lef70,Tieger84,Ananin85,belli99} and with theoretical 
predictions \cite{drechsel99}. Apparently, the present data are of 
considerable superiority to the previous ones, both with respect to  
self-consistency and with respect to completeness. Where available, the BNL data
\cite{belli99}    are in a general reasonable agreement with the 
present ones except for some deviations especially at the highest 
energies accessible in  \cite{belli99}.
The experimental  data 
are compared with predictions  \cite{drechsel99} based on an extended version  
of the distorted wave impulse approximation (DWIA) 
\cite{Chu87,Gmi87,Kamalov85,Kamalov97}.   
In this approach the elementary process is described by the
recently developed unitary isobar model   \cite{drechsel99a}. 
Medium effects are considered by introducing a phenomenological
$\Delta$ self-energy. This allows to incorporate these effects in a
self-consistent way with regard to both the ``bare''   $\gamma N\Delta$
vertices and $\Delta$ excitations due to pion rescattering.

Without going into details here, the following points should be 
discussed:
In Figs. \ref{fig3} and \ref{fig4} where the differential cross 
sections are shown, the 
dashed curves represent the results of the distorted wave 
impulse-approximation (DWIA). These  predictions apparently are
not in agreement with the experimental data especially for energies 
not too far away from the maximum of the $\Delta$ resonance.  
The inclusion of  $\Delta$-nuclear interaction effects 
(or $\Delta$ self energy as introduced in \cite{drechsel99})
in the free $\Delta$ propagator
leads to an almost perfect agreement between data and predictions
showing that medium modifications of the main resonance 
characteristics
(width and position)  supposedly take care of that
part of the reaction mechanism which is missing in the DWIA.
To give a quantitative information about the $\Delta$-nuclear interaction 
we quote from ref. \cite{drechsel99} the $\Delta$ self-energy at 
E$_\gamma$ = 290 MeV, being V $= 19 - i 33$ MeV which corresponds to an 
increase  of the $\Delta$ resonance mass and width by 19 and 66 MeV, 
respectively. This result is in reasonable agreement with the results 
obtained in pion-nuclear scattering.
It is interesting to note that by using the same parameters
for the $\Delta$ self energy also the data for coherent $\pi^\circ$ 
photoproduction on $^{12}$C \cite{Arends83,Schmitz96} can be 
described. This supports our conclusion about the general validity 
of the theoretical approach suggested in Ref.~\cite{drechsel99}.

\begin{figure}
\centering\psfig{figure=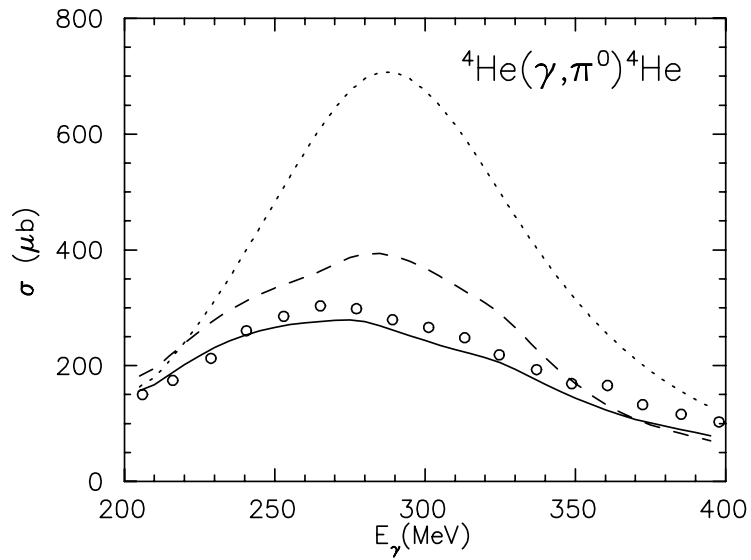,width=.7\linewidth}
\caption{Energy dependence of the total unpolarized cross section for the 
$^4$He$(\gamma,\pi^\circ)^4$He reaction. The experimental data are 
from the present experiment. The dotted and dashed curves are 
the total cross sections for PWIA and DWIA, respectively, of ref.
\cite{drechsel99}. 
The solid curve is the result obtained with the formfactor-type 
parametrization for the $\Delta$ self-energy of ref. \cite{drechsel99}}.
\label{fig5}
\end{figure}

Fig. \ref{fig5} shows the experimental total  cross section for 
coherent $\pi^\circ$
photoproduction obtained in the present experiment 
together with predictions \cite{drechsel99}.  As a dotted line 
this figure also contains the plane-wave impulse approximation 
(PWIA) which was omitted in Figs. \ref{fig3} and \ref{fig4} because 
of its large deviations from the experimental data.
The comparison of the PWIA with the DWIA (dashed curve) shows that
the distortion of the pion wave-function through a pion-nucleus 
optical potential takes care of the largest part of the discrepancy
between experiment and prediction. The final adjustment, then, is
achieved by the introduction of the
medium effects in the $\Delta$ propagator~\cite{drechsel99}.

\subsection{Degrees of linear polarization}

\begin{figure}
\centering\epsfig{figure=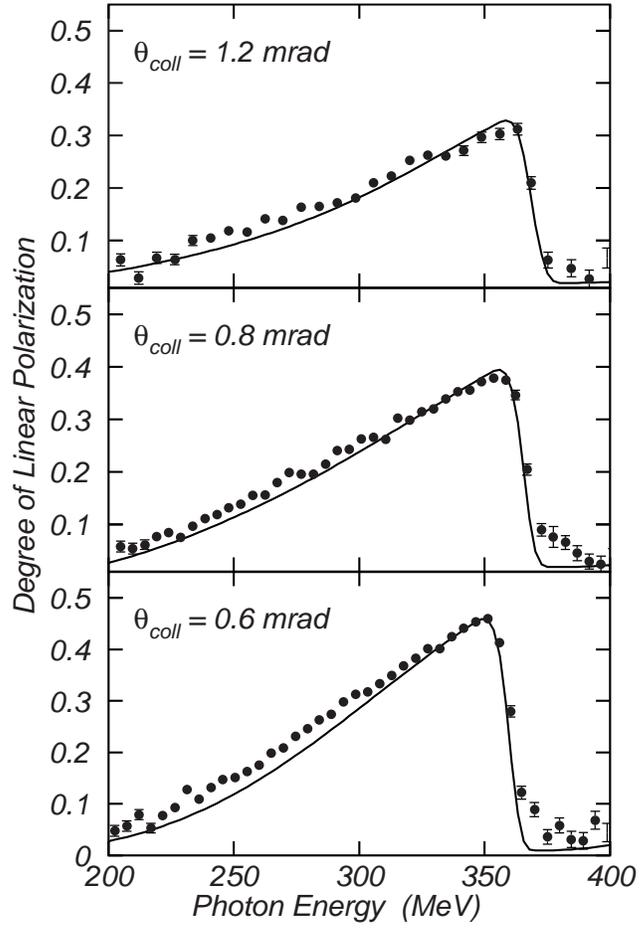,width=.7\linewidth}
\caption{The degree of linear polarization of collimated coherent
bremsstrahlung using collimation angles (half of the angular 
opening) of $\theta_{coll}$ = 1.2, 0.8 and 0.6
mrad. The solid line is the result of a
simulation using the method described in \cite{rambo98}.}
\label{fig6}
\end{figure}

The beam asymmetry measurements have been used to calculate the 
degree of linear polarization of collimated coherent 
bremsstrahlung. The experimental results are shown in Fig. 
\ref{fig6} together with predictions, the
latter ones calculated on an absolute scale 
by the method described in detail in \cite{rambo98}.
Above 300 MeV, i.e. close to the maximum polarization,
the calculation fits the
measured data precisely, whereas below 300 MeV the calculation 
underestimates the measured degree of linear polarization by 
up to 0.03 (absolute value).  This effect appears to increase with 
decreasing collimation angle and is especially pronounced for 
$\theta_{coll}=$ 0.6 mrad. Here we find that the predictions 
systematically  underestimate the data  in the whole range between 
200 and 300 MeV.
This effect had already been observed in our previous investigation
\cite{kraus97} but was to some extent masked by the lower statistical
precision achieved in that experiment for the smallest collimation 
angle. An elaborate investigation carried out in the present work 
showed that this systematic deviation cannot be explained by 
geometrical imperfections of the collimation system, i.e. it 
can neither be explained by a slightly decentral placement
of the collimator nor by effects related with the  beam properties.

It is desirable to clear up the physical origin of the remaining 
discrepancies between predictions and experimental data for the 
degree of linear polarization. Nevertheless, from the point of 
view of the application of collimated coherent 
bremsstrahlung as a source of linearly polarized photons the 
achieved precision is quite satisfactory. The argument for this is 
that in the region of large linear polarization, which is the most 
important one for practical applications, 
the agreement between experiment and calculation is perfect. 
The measurement of linear polarization may be improved by on-line 
monitoring  of the intensity spectrum of coherent bremsstrahlung 
and, furthermore, by developing
reliable methods to convert the intensity spectrum directly into
that of linear polarization.

\section{Summary and conclusion}

Differential cross sections for coherent $\pi^\circ$ 
photoproduction on $^4$He in the energy range between 200 MeV and 
400 MeV  were measured with very high statistical and systematic 
precisions. The experimental data are compared with predictions 
obtained on the basis of the DWIA. These predictions are in perfect 
agreement with the experimental data if appropriate medium 
modifications of the $\Delta(1232)$-resonance width and position are
taken into account through a semiempirical procedure.

Making use of the properties of coherent $\pi^\circ$ 
photoproduction as a polarimeter, the degree  of linear 
polarization of coherent bremsstrahlung has been measured as a 
function of the collimation angle $\theta_{coll}$ of the photons. 
Satisfactory results are obtained with respect to the utilization 
of coherent bremsstrahlung as a source 
of linearly polarized photons.  However, there remain minor 
deviations of the measured polarization data from predicted ones 
which need to be clarified.

 \clearpage

\clearpage


\begin{table}
\caption{Unpolarized differential cross sections (in the $cm$ system) 
for coherent $\pi^{\circ}$ photoproduction
on $^4$He in units of $\mu$b/sr. First column: $\vartheta_{cm}$ production
angle in degrees. Following columns: cross sections with 1$\sigma$ 
statistical errors for the intervals of photon laboratory energy  
(in MeV) given on top of each column. \newline}
\begin{tabular}{c@{\hspace{0.5cm}}cccccc}
$\vartheta_{cm}$ & 201-210 & 211-222 & 223-234 & 235-246 & 247-258 
& 259-270 \\
\hline
7 & 0.1 (0.1) & 0.3 (0.1) & 0.4 (0.1) & 0.4 (0.2) & 1.8 (0.3) 
& 1.3 (0.3)  \\
12 & 2.1 (0.1) & 2.9 (0.1) & 3.6 (0.2) & 5.4 (0.3) & 6.2 (0.3) 
& 7.5 (0.3)  \\
17 & 4.2 (0.2) & 5.5 (0.2) & 6.9 (0.2) & 9.2 (0.3) & 11.2 (0.3) 
& 13.2 (0.3)\\
22 & 6.9 (0.2) & 8.3 (0.2) & 10.6 (0.2) & 14.3 (0.3) & 16.8 (0.3) 
& 20.3 (0.3) \\
27 & 9.2 (0.3) & 11.5 (0.3) & 14.0 (0.3) & 18.6 (0.4) & 22.4 (0.4) 
& 26.6 (0.4) \\
32 & 11.4 (0.3) & 13.3 (0.3) & 15.8 (0.4) & 22.5 (0.4) & 28.7 (0.4) 
& 33.2 (0.4) \\
37 & 12.7 (0.4) & 15.9 (0.4) & 18.9 (0.4) & 26.4 (0.5) & 33.2 (0.5) 
& 39.7 (0.6) \\
42 & 15.0 (0.4) & 17.7 (0.4) & 23.6 (0.5) & 29.9 (0.6) & 37.6 (0.7) 
& 43.2 (0.7) \\
47 & 15.9 (0.4) &     --     &     --     &    --      &     --     
& -- \\
62 &    --      &     --     & 31.5 (0.7) & 41.4 (0.8) & 48.0 (0.8) 
& 54.4 (0.8) \\
67 &    --      & 24.7 (0.7) & 34.8 (0.6) & 40.7 (0.5) & 46.2 (0.5) 
& 49.2 (0.5) \\
72 & 19.5 (0.6) & 25.7 (0.6) & 32.5 (0.5) & 39.0 (0.5) & 42.7 (0.4) 
& 45.3 (0.4) \\
77 & 20.4 (0.5) & 24.4 (0.4) & 30.2 (0.4) & 36.5 (0.4) & 39.5 (0.4) 
& 41.0 (0.4) \\
82 & 19.2 (0.4) & 23.8 (0.4) & 27.5 (0.4) & 32.7 (0.4) & 35.2 (0.4) 
& 35.8 0.3) \\
87 & 17.6 (0.3) & 20.1 (0.3) & 24.9 (0.3) & 29.1 (0.4) & 30.8 (0.4) 
& 30.3 (0.3) \\
92 & 17.0 (0.3) & 19.1 (0.3) & 22.0 (0.3) & 26.0 (0.4) & 27.4 (0.4) 
& 26.6 (0.4) \\
97 & 14.9 (0.3) & 16.7 (0.3) & 19.3 (0.3) & 21.9 (0.4) & 22.5 (0.4) 
& 21.7 (0.5) \\ 
102 & 13.3 (0.2) & 14.1 (0.2) & 15.6 (0.3) & 17.1 (0.3) & 16.4 (0.4) 
& 15.4 (0.5) \\
107 & 11.9 (0.2) & 12.7 (0.2) & 13.0 (0.3) & 14.3 (0.3) & 12.6 (0.4) 
& 12.0 (0.4) \\
112 & 10.3 (0.2) & 11.1 (0.2) & 11.8 (0.3) & 11.9 (0.3) & 10.2 (0.3) 
& 8.9 (0.3) \\
117 & 8.7 (0.2) & 8.8 (0.2) & 9.2 (0.2) & 9.6 (0.2) & 8.2 (0.2) 
& 6.3 (0.2) \\
122 & 7.0 (0.2) & 7.8 (0.2) & 7.4 (0.2) & 7.0 (0.2) & 6.6 (0.2) 
& 4.8 (0.2) \\
127 & 6.1 (0.3) & 6.0 (0.2) & 6.1 (0.3) & 6.4 (0.2) & 5.0 (0.2) 
& 3.6 (0.2) \\
132 & 5.1 (0.3) & 5.1 (0.2) & 5.5 (0.3) & 4.9 (0.3) & 3.7 (0.2) 
& 2.5 (0.1) \\
137 & 3.3 (0.2) & 3.9 (0.2) & 3.1 (0.3) & 3.2 (0.3) & 2.9 (0.2) 
& 1.9 (0.2) \\
142 & 3.1 (0.2) & 3.5 (0.2) & 2.8 (0.4) & 0.8 (0.3) & 1.0 (0.2) 
& 0.5 (0.2) \\
147 & 2.8 (0.2) & 2.3 (0.2) & 1.5 (0.3) &   --      &    --     
&   --      \\

\end{tabular}
\label{sigdif}
\end{table}

\setcounter{table}{0}
\begin{table}
 \caption{(continued)
           \newline}
\begin{tabular}{c@{\hspace{0.5cm}}cccccc}
$\vartheta_{cm}$ & 271-282 & 283-294 & 295-308 & 309-318 & 319-330 
& 331-342 \\
\hline
7 & 2.2 (0.3) & 3.4 (0.3) & 2.9 (0.2) & 4.5 (0.3) & 4.2 (0.3) 
& 2.5 (0.3) \\
12 & 8.8 (0.3) & 10.3 (0.3) & 10.3 (0.3) & 11.9 (0.4) & 11.6 (0.3) 
& 10.9 (0.3) \\
17 & 15.5 (0.3) & 16.7 (0.3) & 19.0 (0.3) & 20.2 (0.4) & 20.5 (0.3) 
& 20.5 (0.3) \\
22 & 23.0 (0.4) & 24.7 (0.4) & 27.3 (0.4) & 29.2 (0.5) & 29.3 (0.4) 
& 30.7 (0.4) \\
27 & 30.0 (0.4) & 32.6 (0.4) & 35.6 (0.4) & 38.4 (0.5) & 37.9 (0.4) 
& 37.4 (0.4) \\
32 & 36.9 (0.4) & 40.3 (0.4) & 42.7 (0.4) & 46.3 (0.6) & 44.6 (0.4) 
& 43.8 (0.4) \\
37 & 44.1 (0.5) & 45.9 (0.5) & 49.3 (0.5) & 50.5 (0.6) & 50.0 (0.5) 
& 47.8 (0.5) \\
42 & 47.7 (0.7) & 50.6 (0.6) & 53.0 (0.6) & 54.4 (0.8) & 51.9 (0.6) 
& 50.0 (0.5) \\
47 & 52.2 (1.2) & 52.6 (1.0) & 54.2 (0.9) & 53.5 (1.0) & 51.7 (0.7) 
& 48.5 (0.7) \\
51 & 51.4 (3.3) & 49.2 (2.4) & 55.4 (2.0) & 55.8 (1.9) & 49.6 (1.2) 
& 45.9 (1.1) \\
58 & 50.0 (1.6) & 51.3 (1.4) & 51.2 (1.4) & 52.3 (1.4) & 48.8 (1.0) 
& 40.7 (0.8) \\
62 & 52.9 (0.7) & 49.7 (0.6) & 50.7 (0.6) & 44.9 (0.7) & 41.0 (0.5) 
& 33.4 (0.4) \\
67 & 49.3 (0.5) & 46.2 (0.4) & 44.2 (0.4) & 39.0 (0.5) & 33.9 (0.3) 
& 28.3 (0.3) \\
72 & 44.0 (0.4) & 41.0 (0.4) & 37.8 (0.3) & 32.9 (0.4) & 27.9 (0.3) 
& 22.0 (0.2) \\
77 & 39.4 (0.3) & 34.4 (0.3) & 31.0 (0.3) & 25.6 (0.3) & 20.8 (0.2) 
& 16.4 (0.2) \\
82 & 33.9 (0.3) & 29.3 (0.3) & 24.7 (0.2) & 20.6 (0.3) & 15.5 (0.2) 
& 12.1 (0.2) \\
87 & 28.2 (0.3) & 23.9 (0.3) & 19.0 (0.2) & 15.0 (0.3) & 10.9 (0.2) 
& 8.8 (0.1) \\
92 & 24.3 (0.4) & 19.0 (0.3) & 14.4 (0.2) & 10.8 (0.3) & 7.8 (0.2) 
& 6.0 (0.1) \\
97 & 20.1 (0.5) & 17.0 (0.4) & 10.8 (0.3) & 8.0 (0.3) & 5.2 (0.2) 
& 4.3 (0.1) \\
101 & 12.1 (0.6) & 10.8 (0.8) & 7.1 (0.5) & 4.9 (0.4) & 3.7 (0.2) 
& 2.1 (0.2) \\
107 & 7.5 (0.5) & 4.7 (0.5) & 1.6 (0.3) & 3.0 (0.4) & 1.9 (0.2) 
& 1.4 (0.2) \\
112 & 6.4 (0.3) & 4.3 (0.2) & 2.6 (0.2) & 2.1 (0.2) & 1.7 (0.1) 
& 1.6 (0.1) \\
117 & 4.1 (0.2) & 2.7 (0.1) & 2.0 (0.1) & 1.8 (0.1) & 1.5 (0.1) 
& 1.3 (0.1) \\
122 & 3.2 (0.1) & 2.0 (0.1) & 1.6 (0.1) & 1.3 (0.1) & 1.4 (0.1) 
& 1.3 (0.1) \\
127 & 2.2 (0.1) & 1.3 (0.1) & 1.3 (0.1) & 1.3 (0.1) & 1.2 (0.1) 
& 1.2 (0.1) \\
132 & 1.4 (0.1) & 0.9 (0.1) & 1.0 (0.1) & 1.1 (0.1) & 1.1 (0.1) 
& 1.0 (0.1) \\
137 & 1.0 (0.1) & 0.8 (0.1) & 0.9 (0.1) & 1.1 (0.1) & 1.1 (0.1) 
& 0.9 (0.1) \\

\end{tabular}
\end{table}

\setcounter{table}{0}
\begin{table}
 \caption{(continued)
 \newline}
\begin{tabular}{c@{\hspace{0.5cm}}ccccc}
$\vartheta_{cm}$ & 343-355 & 356-366 & 367-378 & 379-390 & 391-401 \\
\hline
8 & 2.4 (0.3) & 2.0 (0.4) & -- & 9.9 (3.5) & -- \\
12 & 9.1 (0.3) & 10.7 (0.4) & 5.2 (0.6) & 7.4 (0.8) & 8.4 (1.5) \\
17 & 19.9 (0.3) & 21.4 (0.4) & 18.2 (0.6) & 18.0 (0.7) & 17.0 (0.7) \\
22 & 28.9 (0.4) & 30.9 (0.4) & 27.4 (0.6) & 26.7 (0.6) & 24.2 (0.6) \\
27 & 36.1 (0.4) & 37.7 (0.4) & 34.1 (0.6) & 31.2 (0.5) & 29.2 (0.5) \\
32 & 40.3 (0.4) & 43.3 (0.5) & 37.2 (0.6) & 34.3 (0.5) & 32.3 (0.5) \\
37 & 45.1 (0.4) & 47.0 (0.5) & 40.2 (0.6) & 35.7 (0.6) & 33.1 (0.6) \\
42 & 45.9 (0.5) & 46.5 (0.6) & 38.3 (0.8) & 35.7 (0.8) & 31.1 (0.8) \\
47 & 43.1 (0.7) & 43.8 (0.9) & 35.6 (1.1) & 27.0 (1.1) & 28.7 (1.2) \\
52 & 37.9 (1.0) & 36.8 (1.3) & 27.7 (1.9) & 27.4 (2.2) & 20.3 (2.5) \\
58 & 36.4 (0.8) & 32.4 (1.0) & 20.9 (1.4) & 17.1 (1.8) & 15.0 (2.8) \\
62 & 29.0 (0.4) & 27.1 (0.5) & 19.4 (0.6) & 15.5 (0.7) & 12.6 (0.8) \\
67 & 23.5 (0.3) & 21.9 (0.3) & 16.6 (0.4) & 12.0 (0.4) & 10.3 (0.4) \\
72 & 18.4 (0.2) & 16.7 (0.2) & 12.0 (0.3) & 9.9 (0.3) & 7.4 (0.2) \\
77 & 13.3 (0.2) & 11.8 (0.2) & 8.5 (0.2) & 6.7 (0.2) & 5.4 (0.2) \\
82 & 9.1 (0.1) & 7.9 (0.1) & 5.9 (0.2) & 4.7 (0.2) & 3.7 (0.1) \\
87 & 6.6 (0.1) & 5.9 (0.1) & 4.1 (0.1) & 3.0 (0.1) & 2.6 (0.1) \\
92 & 4.6 (0.1) & 4.0 (0.1) & 2.7 (0.1) & 2.2 (0.1) & 1.5 (0.1) \\
97 & 3.0 (0.1) & 2.8 (0.1) & 1.4 (0.1) & 1.0 (0.1) & 0.8 (0.1) \\
102 & 2.1 (0.1) & 1.9 (0.1) & 0.4 (0.1) & -- & -- \\
107 & 0.2 (0.1) & 0.8 (0.1) & 0.3 (0.1) & -- & -- \\
112 & 1.3 (0.1) & 1.0 (0.1) & 0.8 (0.1) & -- & -- \\
117 & 1.1 (0.1) & 1.2 (0.1) & 0.4 (0.1) & -- & -- \\
122 & 1.1 (0.1) & 1.0 (0.1) & 0.7 (0.1) & -- & -- \\

\end{tabular}
\end{table}

\setcounter{table}{1}
\begin{table}
\caption{Unpolarized total cross section $\sigma_{tot}$ for coherent 
$\pi^{\circ}$ photoproduction 
  on $^4$He in units of $\mu$b as a function of the photon laboratory
  energy $E_{\gamma}$ in MeV. The numbers in parentheses are
  1$\sigma$ statistical errors. \newline}
\begin{tabular}{cc@{\hspace{2cm}}cc}
$E_{\gamma}$ & $\sigma_{tot}$ & $E_{\gamma}$ & $\sigma_{tot}$  \\ 
\hline
206 & 150.0 (0.8) & 313 & 248.1 (0.9) \\
216 & 174.5 (0.8) & 325 & 218.7 (0.6) \\
229 & 212.6 (0.9) & 337 & 193.1 (0.6) \\
241 & 260.3 (1.0) & 349 & 168.9 (0.6) \\
253 & 285.2 (1.0) & 361 & 165.7 (0.7) \\
265 & 303.2 (1.0) & 372 & 132.5 (0.8) \\
277 & 298.3 (0.9) & 385 & 116.0 (0.8) \\
289 & 279.3 (0.8) & 397 & 102.8 (0.8) \\
301 & 266.1 (0.8) 

\end{tabular}
\label{sigtot}
\end{table}

\clearpage


\newpage

\newpage

\newpage

\newpage

\newpage

\newpage

\end{document}